\documentclass [12pt]{article}
\usepackage {amssymb}

\begin {document}

\title {On Weyl  channels being covariant
with respect to the maximum commutative group of unitaries}

\author {Grigori G. Amosov\thanks {E-mail: gramos@mail.ru}\\ Department of Higher Mathematics\\
Moscow Institute of Physics and Technology\\ Dolgoprudny
141700\\RUSSIA}

\maketitle

\abstract {We investigate the Weyl channels being covariant with
respect to the maximum commutative group of unitary operators.
This class includes the quantum depolarizing channel and the
"two-Pauli" channel as well. Then, we show that our estimation of
the output entropy for a tensor product of the phase damping
channel and the identity channel based upon the decreasing
property of the relative entropy allows to prove the additivity
conjecture for the minimal output entropy for the quantum
depolarizing channel in any prime dimesnsion and for the "two
Pauli" channel in the qubit case.}

\section {Introduction}

Let $H$ be a finite-dimensional Hilbert space. Denote $\sigma
(H)$ the set of states (positive unit-trace operators) in $H$.
The linear map $\Phi :\sigma (H)\to \sigma (H)$ is said to be a
quantum channel if $\Phi ^{*}$ is completely positive (\cite
{Hol}). The quantum channel $\Phi $ is called bistochastic if
$\Phi (\frac {1}{d}I_{H})=\frac {1}{d}I_{H}$, where $I_{H}$ is
the identity operator in $H$.

The Holevo-Schumacher-Westmoreland bound $C_{1}(\Phi)$ of a
quantum channel $\Phi $ is defined by the formula
$$
C_1(\Phi )=\sup \limits _{x_j\in \sigma (H),\pi} S(\sum \limits
_{j=1}^r\pi_j\Phi (x_j))-\sum \limits _{j=1}^r\pi_jS(\Phi (x_j)),
$$
where $S(x)=-Trxlogx$ is the von Neumann entropy of $x$ and the
supremum is taken over all probability distributions $\pi
=(\pi_j)_{j=1}^r,\ 0\leq \pi_j\leq 1,\ \sum \limits
_{j=1}^r\pi_j=1$ for all $r\in {\mathbb N}$.  {\it The additivity
conjecture} states (\cite {AHW00}) that for any two channels $\Phi
$ and $\Psi $
$$
C_1(\Phi \otimes \Psi )=C_1(\Phi )+ C_1(\Psi ).
$$
If the additivity conjecture holds, one can easily find the
capacity $C(\Psi )$ of the channel $\Psi $ by the formula $C(\Psi
)=\lim \limits _{n\to +\infty } \frac {C_1(\Psi
^{\otimes n})}{n}=C_1(\Psi )$ (see \cite {H98}).

At the moment, the additivity conjecture is proved for a number of
different cases (\cite {H98, C01, C02, S02, Fan, Dat}). The
crucial role was played by C. King  who proved \cite {C01, C02}
the additivity conjecture for all bistochastic qubit channels as
well as for the quanum depolarizing channel. In \cite {Amo} it
was shown that the additivity conjecture for the channels of the
form $\Psi \circ \Phi $, where $\Phi $ is the quantum
depolarizing channel and $\Psi $ is the phase damping can be
proved by means of the decreasing property of the relative
entropy \cite {Lind}. In the present paper we continue to study a
possibility to use the decreasing property of the relative
entropy to estimate the von Neumann entropy of the output of a
quantum channel.

The additivity conjecture for the quantity $C_{1}$ is closely
connected with the additivity conjecture for the entropy infimum
which states that
$$
\inf \limits _{\rho\in \sigma (H\otimes K)}S((\Phi \otimes
\Psi)(\rho))=\inf \limits _{\rho \in \sigma (H)}S(\Phi (\rho))+
\inf \limits _{\rho \in \sigma (K)}S(\Psi (\rho))
$$
for any two channels $\Phi $ and $\Psi $ acting in Hilbert spaces
$H$ and $K$, respectively. For the case of the Weyl channel we
consider in the present paper, if the additivity conjecture for
the entropy infimum holds it implies that the quantity $C_{1}$ is
additive also (\cite {H05}). As it is shown in \cite {S04} the
additivity conjectures for $C_{1}$ and the entropy infimum are
globally equivalent.

Fix the basis $e_{j},\ 0\le j\le d-1,$ in a Hilbert space $H,\
dimH=d$, and define a unitary operator $U$ by the formula
$$
U=\sum \limits _{j=0}^{d-1}e^{\frac {2\pi i}{d}j}|e_{j}><e_{j}|.
$$
Given a probability distribution $0\le \lambda _{j}\le 1,\ \sum
\limits _{j=0}^{d-1}\lambda _{j}=1,$ we shall call the
bistochastic quantum channel defined as
\begin {equation}\label {PD}
\Phi (x)=\sum \limits _{j=0}^{d-1}\lambda _{j}U^{j}xU^{*j},\ x\in
\sigma (H),
\end {equation}
by {\it a phase damping}. The map
$$
E(x)=\frac {1}{d}\sum \limits _{j=0}^{d-1}U^{j}xU^{*j}\equiv \sum
\limits _{j=0}^{d-1}|e_{j}><e_{j}|x|e_{j}><e_{j}|,
$$
$x\in \sigma (H)$, is a conditional expectation to the algebra of
the elements being fixed with respect to the action of the group
$\{U^{j},\ 0\le j\le d-1\}$ and, hence, of the channel $\Phi $.

Given two states $\rho $ and $\overline \rho $, let $S(\rho
,\overline \rho)=Tr(\rho \log \rho)-Tr(\rho\log \overline \rho)$
be the Umegaki relative entropy \cite {Umegaki}. Due to \cite
{Lind} the following property holds,
$$
S(\Phi (\rho),\Phi (\overline {\rho}))\le S(\rho ,\overline
{\rho})
$$
for any (not necessarily bistochastic) quantum channel $\Phi $.
Using this property it is possible to estimate the entropy of the
output of a quantum channel (\cite {Amo}).

Let $K$ be an arbitary finite-dimensional Hilbert space. Using
the decreasing property of the relative entropy we shall prove
the following statement.

{\bf Theorem 1.}{\it Let $x\in \sigma (H\otimes K)$ be such that
$$
E(Tr_{K}(x))=\frac {1}{d}I_{H}.
$$
Then, for the phase damping (\ref {PD}) we get
\begin {equation}\label {Est}
S((\Phi \otimes Id)(x))\ge -\sum \limits _{j=0}^{d-1}\lambda
_{j}\log \lambda _{j}+\frac {1}{d}\sum \limits
_{j=0}^{d-1}S(x_{j}),
\end {equation}
where $x_{j}=dTr_{H}((|e_{j}><e_{j}|\otimes I_{K})x)\in \sigma
(K),\ 0\le j\le d-1$. }

Analyzing the Weyl channels being covariant with respect to the
maximum commutative group of unitary operators we shall prove the
following two theorems based upon the estimation (\ref {Est}).

{\bf Theorem 2.}{\it Let $\Phi (\rho)=(1-p)\rho+\frac
{p}{d}I_{H},\ \rho\in \sigma (H),\ 0\le p\le \frac
{d^{2}}{d^{2}-1},$ be the quantum depolarizing channel in the
Hilbert space $H$ of the prime dimension $d$. Then, there exists
$d$ orthonormal bases $\{f_{j}^{s},\ 0\le s,j\le d-1\}$ in $H$
such that
\begin {equation}\label {XJ}
S((\Phi \otimes Id)(x))\ge -(1-\frac {d-1}{d}p)\log (1-\frac
{d-1}{d}p)-\frac {d-1}{d}p\log \frac {p}{d}+
\end {equation}
$$
\frac {1}{d^{2}}\sum \limits _{j=0}^{d-1}\sum \limits
_{s=0}^{d-1}S(x_{j}^{s}),
$$
where $x\in \sigma (H\otimes K),\
x_{j}^{s}=dTr_{H}((|e_{j}^{s}><e_{j}^{s}|\otimes I_{K})x)\in
\sigma (K),\ 0\le j,s\le d-1$. }

Denote $\sigma _{x},\sigma _{y}$ and $\sigma _{z}$ the Pauli
operators.

{\bf Theorem 3.}{\it Let $\Phi (\rho)=(1-2p)\rho+p\sigma
_{y}\rho\sigma _{y}+p\sigma _{z}\rho \sigma _{z}$ be the "two
Pauli" channel in the space $H,\ dimH=2$. Suppose that $p\le \frac
{1}{3}$, then there exist three orthonormal bases
$(e_{1}^{s},e_{2}^{s}),\ 1\le s\le 3,$ in $H$ such that
$$
S((\Phi \otimes Id)(x))\ge -(1-p)\log (1-p)-p\log p+ \frac
{1}{6}\sum \limits _{n=1}^{3}(S(x_{j}^{1})+S(x_{j}^{2})),
$$
where $x\in \sigma (H\otimes K),\
x_{j}^{k}=2Tr_{H}((|e_{k}^{j}><e_{k}^{j}|\otimes I_{K})x)\in
\sigma (K),\ 1\le j\le 3,\ 1\le k\le 2$. }

{\bf Remark.}{\it The additivity of the minimal output entropy
for the quantum depolarizing channel and the "two-Pauli" channel
follows from Theorems 2 and 3. Indeed, these theorems contain more
information. It was shown in \cite {C01} that if the additivity
conjecture holds for the qubit depolarizing channel and the
"two-Pauli" channel in the case $p\le \frac {1}{3}$, then it holds
for all bistochastic qubit channels. Hence, in particular,
Theorems 2 and 3 imply that the additivity conjecture holds for
all bistochastic qubit channels.}

This paper is organized as follows. In Part 2 we prove Theorem 1.
Part 3 is devoted to the description of the Weyl (the same as
Weyl-covariant) channels. The orbits of the maximum commutative
groups in the dimensions $d=2$ and $d=3$ are described in Part 3.
In Part 4 we find the conditions which are sufficient for the
Weyl channel be covariant with respect to the maximum commutative
group.  In Part 5 the proof of Theorems 2 and 3 is given.

\section {Proof of Theorem 1}

In this section we shall use the approach introduced in \cite
{Amo}.

Fix $x$ with the property formulated in Theorem 1 and define a
quantum channel $\Xi _{x}:\sigma (H\otimes K)\to \sigma (H\otimes
K)$ by the formula
$$
\Xi _{x}(\rho)=\sum \limits _{j=0}^{d-1}Tr((|e_{j}><e_{j}|\otimes
I_{K})\rho )(U^{j}\otimes I_{K})x(U^{*j}\otimes I_{K}),\ \rho \in
\sigma (H\otimes K).
$$
Put $\rho =\sum \limits _{j=0}^{d-1}\lambda
_{j}|e_{j}><e_{j}|\otimes y,\ \overline \rho =\sum \limits
_{j=0}^{d-1}\frac {1}{d}|e_{j}><e_{j}|\otimes y=\frac
{1}{d}I_{H}\otimes y$, where $y\in \sigma (K)$ is an arbitrary
fixed state. Then,
$$
\Xi _{x}(\rho)=(\Phi \otimes Id)(x),
$$
$$
\Xi _{x}(\overline \rho)=\frac {1}{d}\sum \limits
_{j=0}^{d-1}(U^{j}\otimes I_{K})x(U^{*j}\otimes I_{K})\equiv
\tilde E(x).
$$
Notice that $\tilde E=(E\otimes Id)$ is the conditional
expectation to algebra of the elements being fixed with respect to
the action of the cyclic group $\{U^{j}\otimes I_{K} ,\ 0\le j\le
d-1 \}$.

The decreasing property of the relative entropy \cite {Lind} gives
us the estimation:
\begin {equation}\label {EE2}
S(\Xi_{x}(\rho),\Xi _{x}(\overline \rho))\le S(\rho ,\overline
\rho)=\sum \limits _{j=0}^{d-1}\lambda _{j}\log\lambda
_{j}+\log(d).
\end {equation}

On the other hand,
$$
S(\Xi _{x}(\rho),\Xi _{x}(\overline \rho))=Tr((\Phi \otimes
Id)(x)\log(\Phi \otimes Id)(x))-
$$
$$
Tr((\Phi \otimes Id)(x)\log \tilde E(x))=-S((\Phi \otimes Id)(x))-
$$
$$
Tr(\tilde E\circ (\Phi \otimes Id)(x)\log \tilde E(x))=
$$
\begin {equation}\label {EE1}
-S((\Phi \otimes Id)(x))+S(\tilde E(x)).
\end {equation}
Here we used the equality $\tilde E\circ (\Phi \otimes Id)=\tilde
E$ which holds because $\tilde E$ is the conditional expectation
to the algebra of elements being fixed with respect to the action
of $\Phi \otimes Id$. It follows from the condition of Theorem 1
that
$$
\tilde E(x)=\frac {1}{d}\sum \limits
_{j=0}^{d-1}|e_{j}><e_{j}|\otimes x_{j},\ x_{j}\in \sigma (K).
$$
Thus,
\begin {equation}\label {EE3}
S(\tilde E(x))=log(d)+\frac {1}{d}\sum \limits
_{j=0}^{d-1}S(x_{j}),
\end {equation}
$$
$$
$x_{j}=dTr_{H}((|e_{j}><e_{j}|\otimes I_{K})x),\ 0\le j\le d-1$.
Combining (\ref {EE2}), (\ref {EE1}) and (\ref {EE3}) we get the
result we need.

\section {The Weyl channels}

Recently the representation of quantum channels by means of the
discrete Weyl group were discussed in different contexts (\cite
{Amo, Ruskai, Fukuda, Cerf}). Fix the orthonormal basis $|k>,\
k=0,1,\dots ,d-1$ of the Hilbert space $H,\ dimH=d,$ and define
the unitary operators $U_{m,n}$ by the formula (\cite {Fivel})
$$
U_{m,n}=\sum \limits _{k=0}^{d-1}e^{\frac {2\pi i}{d}kn}|k+m\
mod\ d
><k|,
$$
$0\le m,n\le d-1$. The operators $U_{m,n}$ are satisfied the Weyl
commutation relations,
\begin {equation}\label {WO}
U_{m,n}U_{m',n'}=e^{2\pi i(m'n-mn')/d}U_{m',n'}U_{m,n},
\end {equation}
$0\le m,n,m',n'\le d-1$. We shall call $U_{m,n}$ by a Weyl
operator. Notice that
\begin {equation}\label {WOA}
U_{m,0}|k>=|k+m\ mod\ d>,\ U_{0,n}|k>=e^{\frac {2\pi i}{d}kn\ mod\
d}|k>,
\end {equation}
$$
U_{m,n}=U_{m}U_{n},\ 0\le m,n\le d-1.
$$
We shall consider bistochastic quantum communication channels of
the following form
\begin {equation}\label {Weyl}
\Phi (x)=\sum \limits _{m,n=0}^{d-1}\pi _{m,n}U_{m,n}xU_{m,n}^{*},
\end {equation}
$x\in \sigma (H)$, where $0\le \pi _{m,n}\le 1,\ \sum \limits
_{m,n=0}^{d-1}\pi _{m,n}=1,$ is an arbitrary probability
distribution. We shall call (\ref {Weyl}) by a Weyl channel.

The bases $(e_{j})_{j=0}^{d-1}$ and $(f_{j})_{j=0}^{d-1}$ of the
space $H$ are said to be mutually unbiased bases (\cite {Ivan}) if
$|<e_{j}|f_{k}>|=\frac {1}{\sqrt d},\ 0\le j,k\le d-1$. For
example, the eigenvectors $f_{j}=|j>$ and $e_{j}=\frac {1}{\sqrt
d}\sum \limits _{k=0}^{d-1}e^{\frac {2\pi i}{d}jk}|k>$ of the
Weyl operators $U_{0,n}$ and $U_{m,0}$, respectively, form
mutually unbiased bases. If the dimension $d$ of the Hilbert
space $H$ is prime, there exist $d+1$ mutually unbiased bases
$\{e_{j}^{k},\ 0\le j\le d-1,\ 0\le k\le d\}$ which can defined as
the eigenvectors of the Weyl operators $U_{sk,k}$ such that
\begin {equation}\label {MUB1}
|e_{j}^{s}><e_{j}^{s}|=\frac {1}{d}\sum \limits
_{k=0}^{d-1}e^{\frac {2\pi i}{d}jk}U_{sk,k},\ 0\le s\le d-1,
\end {equation}
$$
|e_{j}^{d}><e_{j}^{d}|=\frac {1}{d}\sum \limits
_{k=0}^{d-1}e^{\frac {2\pi i}{d}jk}U_{k,0}\equiv |j><j|.
$$
Because the elements of the mutually unbiased bases $(e_{j}^{s})$
are eigenvectors for the Weyl operators $U_{sk,k}$ and $U_{k,0}$,
we get
$$
U_{sk,k}=\sum \limits _{j=0}^{d-1}e^{\frac {2\pi i
}{d}jk}|e_{j}^{s}><e_{j}^{s}|,\ 0\le s\le d-1,
$$
$$
U_{k,0}=\sum \limits _{j=0}^{d-1}e^{\frac {2\pi
i}{d}jk}|e_{j}^{d}><e_{j}^{d}|.
$$

Fix a basis $(e_{j})_{j=0}^{d-1}$ in the space $H$. We shall call
by the maximimum commutative group of unitaries ${\mathcal
U}_{d}((e_{j})_{j=0}^{d-1})$ a $d$-parameter group of unitary
operators
\begin {equation}\label {G}
U=\sum \limits _{j=0}^{d-1}e^{i\phi _{j}}|e_{j}><e_{j}|,
\end {equation}
where $\phi _{j},\ 0\le j\le d-1,$ are arbitrary real numbers.
Denote by ${\mathcal U}_{d}^{s}={\mathcal
U}_{d}((e_{j}^{s})_{j=0}^{d-1}),\ 0\le s\le d,$ the maximum
commutative groups generated by the mutually unbiased bases (\ref
{MUB1}).

The bistochastic quantum channel
$$
E_{s}(x)=\sum \limits
_{j=0}^{d-1}|e_{j}^{s}><e_{j}^{s}|x|e_{j}^{s}><e_{j}^{s}|,\ x\in
\sigma (H),
$$
is a conditional expectation to the algebra of elements being
fixed with respect to the action of the maximum commutative group
${\mathcal U}_{d}^{s}$ as well as of the phase damping channel
$$
\Phi _{s}(x)=\sum \limits _{j=0}^{d-1}\lambda
_{j}U_{sj,j}xU_{sj,j}^{*}
$$
for $0\le s\le d-1$ and
$$
\Phi _{d}(x)=\sum \limits _{j=0}^{d-1}\lambda _{j}U_{j,0}xU_{j,0}
$$
for $s=d$.

\section {Orbits of the maximum commutative group of unitaries.}

Denote by ${\mathcal G}_{d}^{s}$ the set consisting of the unit
vectors $g$ for which there exists the unitary operator $U_{s}\in
{\mathcal U}_{d}^{s},\ 0\le s\le d-1,$ such that
$$
E_{d}(U_{s}|g><g|U_{s}^{*})=\frac {1}{d}I_{H},\
$$
Let ${\mathcal A}_{d}^{s}$ be the convex set of states generated
by the one-dimensional projections $|g><g|,\ g\in {\mathcal
G}_{d}$.

{\bf Lemma 1.}{\it Given complex numbers $a,b,\ |a|^{2}+|b|^{2}=1$
there exist the real numbers $\phi ,\psi,\ \alpha \in [0,2\pi]$
such that
$$
e^{i\phi}a+e^{i\psi}b=1,
$$
$$
e^{i\phi}a-e^{i\psi}b=e^{i\alpha}.
$$
}

{\bf Proof.}

Pick up $\tilde \phi,\tilde \psi$ such that $\cos\tilde
\phi=\sin\tilde \psi=|a|,\ -\sin\tilde \phi=\cos\tilde \psi=|b|$.
Then, $\phi =\tilde \psi -arg(a),\ \psi=\tilde \psi-arg(b),\
\cos\alpha =\cos(2\tilde\phi),\ \sin\alpha =\sin(2\tilde \phi)$
give us the solution. $\Box$

{\bf Proposition 1.} {\it Suppose that $d=2$. Then, ${\mathcal
G}_{2}^{0}={\mathcal G}_{2}^{1}=H$ and, hence, ${\mathcal
A}_{2}^{0}={\mathcal A}_{2}^{1}=\sigma(H)$.}

{\bf Proof.}

We shall prove that ${\mathcal G}_{2}^{0}=H$. The identity
${\mathcal G}_{2}^{1}=H$ can be considered analogously.

Take a vector $g\in H$. Suppose that the coordinates of $g$ in
the basis $e_{0}^{0},e_{1}^{0}$ are $<e_{0}^{0}|g>=a,\
<e_{1}^{0}|g>=b$. Let the numbers $\phi ,\psi ,\alpha \in
[0,2\pi]$ are defined by Lemma 1. Determine the unitary operator
$U\in {\mathcal U}_{2}^{0}$ by the formula
$$
U=e^{i\phi}|e_{0}^{0}><e_{0}^{0}|+e^{i\psi}|e_{1}^{0}><e_{1}^{0}|,
$$
then, it follows from Lemma 1, that
$$
<e_{0}^{2}|Ug>=\frac {1}{\sqrt 2},\ <e_{1}^{2}|Ug>=\frac
{e^{i\alpha }}{\sqrt 2}.
$$
$\Box $

{\bf Proposition 2.}{\it ${\mathcal G}_{3}^{s}\neq H,\ 0\le s\le 2
$.}

{\bf Proof.}

We shall prove that ${\mathcal G}_{3}^{0}\neq H$. The other cases
can be considered analogously.

Take a unit vector $g\in H$ whose coordinates
$<e_{j}^{0}|g>=\alpha _{j},\ 0\le j\le 2,$. If ${\mathcal
G}_{3}^{0}=H$, then there exist the real numbers $\phi _{j},\beta
_{j},\ 0\le j\le 2$, satisfying the relations
\begin {equation}\label {Hren}
\sum \limits _{k=0}^{2}e^{i\phi _{k}}e^{i\frac
{2\pi}{3}jk}=e^{i\beta _{j}},\ 0\le j\le 2.
\end {equation}
Suppose that $\alpha _{1},\alpha _{2}>0$ and $\alpha _{3}=0$, then
the condition (\ref {Hren}) implies that
$$
\cos (\phi _{1}-\phi _{2})=0,
$$
$$
\cos (\phi _{1}-\phi _{2}-\frac {2\pi}{3})=0
$$
and
$$
\cos (\phi _{1}-\phi _{2}-\frac {4\pi}{3})=0.
$$
But the system of these three equation has no solution. $\Box $

{\bf Lemma 2.}{\it Given complex numbers $\alpha _{j},\ \sum
\limits _{j=0}^{2}|\alpha _{j}|^{2}=1$ satisfying the relations
$$
|\alpha _{1}\alpha _{2}|+|\alpha _{1}\alpha _{3}|>|\alpha
_{2}\alpha _{3}|,
$$
$$
|\alpha _{1}\alpha _{2}|+|\alpha _{2}\alpha _{3}|>|\alpha
_{1}\alpha _{3}|,
$$
$$
|\alpha _{1}\alpha _{3}|+|\alpha _{2}\alpha _{3}|>|\alpha
_{1}\alpha _{2}|.
$$
there exist the real numbers $\phi _{j},\beta _{j},\ 0\le j\le
2,$ such that
$$
\sum \limits _{j=0}^{2}e^{i(\phi _{j}+\frac {2\pi jk}{3})}\alpha
_{j}=e^{i\beta _{k}},\ 0\le k\le 2.
$$
}

{\bf Proof.}

Because $\alpha _{j}=|\alpha _{j}|e^{iarg(\alpha _{j})}$, it
suffices to consider only the case of non-negative real numbers
$\alpha _{j}\ge 0,\ 0\le j\le 2$. Denote by $\gamma _{j},\ 0\le
j\le 2,$ the angles of the triangle with the sides equal to the
values $\alpha _{1}\alpha _{2},\ \alpha _{1}\alpha _{3}$ and
$\alpha _{2}\alpha _{3}$. Then,
\begin {equation}\label {F1}
\frac {\sin \gamma _{1}}{\alpha _{1}\alpha _{2}}=\frac {\sin
\gamma _{2}}{\alpha _{1}\alpha _{3}}=\frac {\sin \gamma
_{3}}{\alpha _{2}\alpha _{3}},
\end {equation}
$$
\alpha _{1}\alpha _{2}=\alpha _{1}\alpha _{3}\cos(\gamma
_{3})+\alpha _{2}\alpha _{3}\cos (\gamma _{2}),
$$
\begin {equation}\label {F2}
\alpha _{1}\alpha _{3}=\alpha _{1}\alpha _{2}\cos (\gamma
_{3})+\alpha _{2}\alpha _{3}\cos (\gamma _{1}),
\end {equation}
$$
\alpha _{2}\alpha _{3}=\alpha _{1}\alpha _{2}\cos (\gamma
_{2})+\alpha _{1}\alpha _{3}\cos (\gamma _{1}).
$$
Put
$$
\phi _{1}=0,\ \phi _{2}=\frac {2\pi}{3}+\frac {1}{3}\gamma
_{2}-\frac {1}{3}\gamma _{3},
$$
$$
\phi _{3}=\frac {\pi}{3}-\frac {1}{3}\gamma _{2}-\frac
{2}{3}\gamma _{3}.
$$
Thus, we get
$$
\alpha _{1}\alpha _{2}\cos(\phi _{1}-\phi _{2})+\alpha _{1}\alpha
_{3}\cos (\phi _{1}-\phi _{3})+\alpha _{2}\alpha _{3}\cos (\phi
_{2}-\phi _{3})=
$$
$$
\alpha _{1}\alpha _{2}\cos (\phi _{2})-\alpha _{1}\alpha _{3}\cos
(\phi _{2}+\gamma _{3})-\alpha _{2}\alpha _{3}\cos (\phi
_{2}-\gamma _{2})=
$$
\begin {equation}\label {F3}
(\alpha _{1}\alpha _{2}-\alpha _{1}\alpha _{3}\cos (\gamma
_{3})-\alpha _{2}\alpha _{3}\cos (\gamma _{2}))\cos (\phi _{2})+
\end {equation}
$$
(\alpha _{1}\alpha _{3}\sin (\gamma _{3})-\alpha _{2}\alpha
_{3}\sin (\gamma _{2}))\sin (\phi _{2})=0
$$
due to (\ref {F1}) and (\ref {F2}). Analogously,
$$
\alpha _{1}\alpha _{2}\sin (\phi _{1}-\phi _{2})-\alpha
_{1}\alpha _{3}\sin (\phi _{1}-\phi _{3})+\alpha _{2}\alpha
_{3}\sin (\phi _{2}-\phi _{3})=
$$
$$
-\alpha _{1}\alpha _{2}\sin \phi _{2}+\alpha _{1}\alpha
_{3}\sin(\phi _{2}+\gamma _{3})+\alpha _{2}\alpha _{3}\sin (\phi
_{2}-\gamma _{2})=
$$
\begin {equation}\label {F4}
(-\alpha _{1}\alpha _{2}+\alpha _{1}\alpha _{3}\cos (\gamma
_{3})+\alpha _{2}\alpha _{3}\cos (\gamma _{2}))\sin (\phi _{2})+
\end {equation}
$$
(\alpha _{1}\alpha _{3}\sin(\gamma _{3})-\alpha _{2}\alpha
_{3}\sin (\gamma _{2}))\cos (\phi _{2})=0.
$$
Denote
$$
I_{k}=\sum \limits _{j=0}^{2}e^{i(\phi _{j}+\frac {2\pi
jk}{3})}\alpha _{j},\ 0\le k\le 2.
$$
Under the condition $\sum \limits _{j=0}^{2}\alpha _{j}^{2}=1$
the formulas (\ref {F3})-(\ref {F4}) imply that
$$
Re(I_{k})^{2}+Im(I_{k})^{2}=1,\ 0\le k\le 2.
$$
$\Box $

{\bf Proposition 3.} {\it The set ${\mathcal G}_{3}^{s}$ includes
the unit vectors $h$ whose coordinates $\alpha _{1},\alpha _{2}$
and $\alpha _{3}$ in the basis $(e_{j}^{s})$ satisfy the relation
$$
|\alpha _{1}\alpha _{2}|+|\alpha _{1}\alpha _{3}|>|\alpha
_{2}\alpha _{3}|,
$$
\begin {equation}\label {Triang}
|\alpha _{1}\alpha _{2}|+|\alpha _{2}\alpha _{3}|>|\alpha
_{1}\alpha _{3}|,
\end {equation}
$$
|\alpha _{1}\alpha _{3}|+|\alpha _{2}\alpha _{3}|>|\alpha
_{1}\alpha _{2}|.
$$
}

{\bf Remark.}{\it We doesn't consider the trivial case $\alpha
_{i}=\delta _{ii_{0}}$ for some $i_{0},0\le i_{0}\le 2$.}

{\bf Proof.}

As it was done above we shall consider only the case $s=0$.

Take a vector $g$ satisfying the relations (\ref {Triang}).
Suppose that the coordinates of $g$ in the basis
$e_{0}^{0},e_{1}^{0},e_{2}^{0}$ are $<e_{k}^{0}|g>=\alpha _{k},\
0\le k\le 2$. Let the numbers $\phi _{j},\ 0\le j\le 2,$ are
defined by Lemma 2. Determine the unitary operator $U\in {\mathcal
U}_{3}^{0}$ by the formula
$$
U=\sum \limits _{j=0}^{2}e^{i\phi _{j} }|e_{j}^{0}><e_{j}^{0}|,
$$
then, it follows from Lemma 2 that
$$
|<e_{j}^{3}|Ug>|=\frac {1}{\sqrt 3},\ 0\le j\le 2.
$$
$\Box $

 Denote by $M_{d}$ and $M(0)$ the algebra of all $d\times d$
 matrices and its subalgebra generated by the cyclic group
$\{U_{m,0},\ 0\le n\le d-1\}$, respectively. Let ${\mathcal
U}_{d}$ be the maximum commutative group constructed by means of
the basis $(e_{j})_{j=0}^{d-1}$.

{\bf Proposition 4.}{\it The inclusion $W\in M(0)$ holds for all
$W\in {\mathcal U}_{d}$.}

{\bf Proof.}

It suffices to notice that $U_{m,0}\in {\mathcal U}_{d},\ 0\le
m\le d-1$.

$\Box $

\section {The covariant Weyl channels.}

The quantum channel $\Phi $ is said to be covariant with respect
to the group $\mathcal U$ being a subgroup of the group of all
unitary operators in $H$ if
$$
\Phi (UxU^{*})=U\Phi (x)U^{*},\ x\in \sigma (H),\ U\in {\mathcal U
}.
$$

The Weyl operators $U_{m,n}$ satisfying the relation (\ref {WO})
form the basis in the algebra of all $d\times d$ matrices.
Moreover, the action of a Weyl channel (\ref {Weyl}) on the Weyl
operator $U_{s,t}$ is given by the formula
$$
\Phi (U_{s,t})=\lambda _{st}U_{s,t},
$$
where
$$
\lambda _{st}=\sum \limits _{m,n=0}^{d-1}\pi _{m,n}e^{2\pi
i(sn-tm)/d}.
$$

{\bf Proposition 5.} {\it Suppose that $\lambda _{st}=\mu
_{t}=const,\ 0\le s\le d-1,\ 1\le t\le d-1$. Then the Weyl channel
is covariant with respect to the maximum commutative group
${\mathcal U}_{d}^{d}$.}

{\bf Remark.} {\it If $d=2$, a role of the maximum commutative
group of unitaries can be played by all multiples of the group
$SO(2)$ consisting of all rotations in $H$ implemented by the
matrices of the form
$$
e^{i\psi }\left (\begin {array}{cc}\cos\phi & \sin\phi \\-\sin
\phi & \cos \phi \end {array}\right ),\ \phi,\psi \in [0,2\pi].
$$
It is straightforward to check that the bistochastic qubit
channel determined by the triple $[\lambda _{1},\lambda
_{2},\lambda _{3}]$ is covariant with respect to the group of
rotations iff $\lambda _{1}=\lambda _{3}$. Thus, this class
includes the quantum depolarizing channel $[\lambda ,\lambda
,\lambda]$ and the "two-Pauli" channel
$[\lambda,1-2\lambda,\lambda ]$ as well.}

{\bf Proof.}

Denote $M(n)$ the algebra of matrices generated by the Weyl
operators $U_{m,n},\ 0\le m\le d-1$. Due to Proposition 4 a
unitary operator $W\in {\mathcal U}_{d}^{d}$ can be represented as
$$
W=\sum \limits _{m=0}^{d-1}c_{m}U_{m,0}.
$$
Hence $WU_{m,0}W^{*}=U_{m,0},\ 0\le m\le d-1,$ and
$WM(n)W^{*}\subset M(n)$ for all $W\in {\mathcal U}_{d}^{d},\ 1\le
n\le d-1$. It follows that if the condition of Proposition 5
holds, we get $WU_{s,t}W^{*}\in M(t)$ for $1\le t\le d-1$. Hence,
$$
\Phi (WU_{s,t}W^{*})=\mu _{t}WU_{st}W^{*}= W\Phi (U_{s,t})W^{*},
$$
$0\le s,t\le d-1$. $\Box $

{\bf Proposition 6.} {\it Suppose that $\pi _{mn}=p_{m},\ 0\le
m\le d-1,\ 1\le n\le d-1$. Then the Weyl channel (\ref {Weyl}) is
covariant with respect to the maximum commutative group
${\mathcal U}_{d}^{d}$.}

{\bf Remark.} {\it If $\pi _{00}=(1-\frac {d^{2}-1}{d^{2}}p)$ and
$\pi _{m0}=p_{n}=\frac {p}{d^{2}},\ 1\le n,m\le d-1$, the Weyl
channel (\ref {Weyl}) is the quantum depolarizing channel,
$$
\Phi (x)=(1-p)x+\frac {p}{d}Tr(x),\ 0\le p\le \frac
{d^{2}}{d^{2}-1}.
$$
 Suppose that $d=2$, then the "two Pauli" channel
\begin {equation}\label {qubit}
\Phi (\rho)=(1-2p)\rho+p\sigma _{y}\rho \sigma _{y}+p\sigma
_{z}\rho \sigma _{z},\ \rho \in \sigma (H),
\end {equation}
is satisfied the conditions of Proposition 6 if we identify the
Weyl operators with the Pauli operators such that $U_{0,0}\equiv
I_{H},\ U_{1,0}\equiv \sigma _{x},\ U_{0,1}\equiv \sigma _{y},\
U_{1,1}\equiv i\sigma _{z}$. }

{\bf Proof.}

In the case,
$$
\lambda _{st}=\sum \limits _{m=0}^{d-1}\pi _{m0}e^{-2\pi
itm/d}+\sum \limits _{n=1}^{d-1}\sum \limits
_{m=0}^{d-1}p_{m}e^{2\pi i(sn-tm)/d}=
$$
$$
\sum \limits _{m=0}^{d-1}(\pi _{m0}-p_{m})e^{-2\pi itm/d}\equiv
\mu _{t},
$$
$0\le s\le d-1,\ 1\le t\le d-1$. So, the result follows from
Proposition 5.

$\Box $

 Fix the positive numbers $0\le p_{n}\le 1,\ 0\le r_{m}\le
1,\ 1\le n\le d-1,\ 0\le m\le d-1,$ such that $d\sum \limits
_{n=1}^{d-1}p_{n}+\sum \limits _{m=0}^{d-1}r_{m}=1$ and consider
the Weyl channel
\begin {equation}\label {Chan}
\Phi (x)=\sum \limits _{m=0}^{d-1}r_{m}U_{m,0}xU_{m,0}^{*}+\sum
\limits _{m=0}^{d-1}\sum \limits _{n=1}^{d-1}p_{n}
U_{m,n}xU_{m,n}^{*},
\end {equation}
$x\in \sigma (H)$.

{\bf Proposition 7.} {\it Suppose that $d$ is a prime number, then
the Weyl channel (\ref {Chan}) can be represented in the form
\begin {equation}\label {Equat}
\Phi (x)=\sum \limits _{k=0}^{d-1}\sum \limits
_{m=0}^{d-1}c_{m}U_{m,0}\Phi _{k}(x)U_{m,0}^{*},
\end {equation}
where
$$
\Phi _{k}(x)=\sum \limits _{n=0}^{d-1}\lambda _{n}U_{nk\ mod\
d,n}xU_{nk\ mod\ d,n}^{*},\ x\in \sigma (H),
$$
are the phase damping channels and
$$
\lambda _{0}=1-d\sum \limits _{n=1}^{d-1}p_{n},\ \lambda
_{n}=dp_{n}, 1\le n\le d-1,
$$
$$
c_{m}=\frac {r_{m}}{d(1-d\sum \limits _{n=1}^{d-1}p_{n})},\ 0\le
m\le d-1.
$$
}

{\bf Remark.} {\it Suppose that $d=2$, then the qubit Weyl
channel (\ref {Chan}) has the following form,
$$
\Phi (\rho)=r_{0}\rho +r_{1}\sigma _{x}\rho \sigma
_{x}+p_{1}\sigma _{y}\rho \sigma _{y}+p_{1}\sigma _{z}\rho \sigma
_{z},\ \rho \in \sigma (H),
$$
where we have identified the Weyl operators with the Pauli
matrices such that $U_{0,0}\equiv I_{H},\ U_{1,0}\equiv \sigma
_{x},\ U_{0,1}\equiv \sigma _{y},\ U_{1,1}\equiv i\sigma _{z}$. In
the case, in the representation (\ref {Equat}) we get
$$
\Phi (\rho)=\frac {r_{0}}{2(1-2p_{1})}(\Phi _{0}(\rho)+\Phi
_{1}(\rho))+\frac {r_{1}}{2(1-2p_{1})}\sigma _{x}(\Phi
_{0}(\rho)+\Phi _{1}(\rho))\sigma _{x},
$$
$$
\Phi _{0}(\rho)=(1-2p_{1})\rho+2p_{1}\sigma _{y}\rho \sigma _{y},
$$
$$
\Phi _{1}(\rho)=(1-2p_{1})\rho+2p_{1}\sigma _{z}\rho \sigma _{z},
$$
$\rho \in \sigma (H)$. }

{\bf Proof.}

Let us compare the equations (\ref {Chan}) and (\ref {Equat}). In
(\ref {Equat}) it is included $d$ terms $c_{0}\lambda _{0}x$, $d$
terms $c_{m}\lambda _{0}U_{m,0}xU_{m,0}^{*}$  as well as $d$ terms
$$
c_{m-nk\ mod\ d}\lambda _{n}U_{m,n}xU_{m,n}^{*},\ n\neq 0,
$$
(corresponding to each the channel $\Phi _{k}$). Hence,
$$
dc_{m}\lambda _{0}=r_{m},\ 0\le m\le d-1,
$$
$$
\sum \limits _{k=0}^{d-1}c_{m-nk\ mod\ d}\lambda _{n}= \sum
\limits _{k=0}^{d-1}c_{k}\lambda _{n} =dp_{n},\ 1\le n\le d-1.
$$
Then, let us claim $\sum \limits _{n=0}^{d-1}\lambda _{n}=\sum
\limits _{n=0}^{d-1}c_{n}=1$. It follows that
$$
dc_{m}\lambda _{0}=r_{m},\ 0\le m\le d-1,
$$
$$
\lambda _{n}=dp_{n},\ 1\le n\le d-1.
$$
Hence,
$$
\lambda _{0}=1-d\sum \limits _{n=1}^{d-1}p_{n},
$$
$$
c_{m}=\frac {r_{m}}{d(1-d\sum \limits _{n=1}^{d-1}p_{n})},\ 0\le
m\le d-1.
$$

$\Box $

\section {The estimation of the output entropy.}

Using Theorem 1 and Propositions 7 we can prove Theorem 2.

{\bf Proof of Theorem 2.}

For $x\in \sigma (H\otimes K)$ choose the unitary operator $W$ in
$H$ such that $Tr_{K}(y)=WTr_{K}(x)W^{*}\in {\mathcal A}_{fix}$,
where $y=(W\otimes I_{K})x(W^{*}\otimes I_{K})$. Here we denoted
${\mathcal A}_{fix}$ the algebra of elements being fixed with
respect to the action of the group ${\mathcal U}_{d}^{d}$. The
algebra ${\mathcal A}_{fix}$ is generated by the projections
$|e_{j}^{d}><e_{j}^{d}|$. Hence,
$$
E_{s}(Tr_{K}(y))=\frac {1}{d}I_{H},\ 0\le s\le d-1,
$$
because the bases $(e_{j}^{s})$ are mutually unbiased. Using the
covariance of $\Phi $ with respect to the group of all unitary
operators in $H$ we get
$$
S((\Phi \otimes Id)(x))=S((\Phi \otimes Id)(y)).
$$

Due to Proposition 7 the tensor product of the quantum
depolarizing channel and the identity channel can be represented
as follows
\begin {equation}\label {Equat1}
(\Phi \otimes Id)(y)=\sum \limits _{k=0}^{d-1}\sum \limits
_{m=0}^{d-1}c_{m}(U_{m,0}\otimes I_{K})(\Phi _{k}\otimes Id)
(y)(U_{m,0}^{*}\otimes I_{K}),
\end {equation}
where
$$
(\Phi _{k}\otimes Id)(y)=\sum \limits _{n=0}^{d-1}\lambda
_{n}(U_{nk\ mod\ d,n}\otimes I_{K})y(U_{nk\ mod\ d,n}^{*}\otimes
I_{K}),
$$
$$
\lambda _{0}=1-\frac {d-1}{d}p,\ \lambda _{n}=\frac {p}{d}, 1\le
n\le d-1,
$$
$$
c_{0}=\frac {1-\frac {d^{2}-1}{d^{2}}p}{d(1-\frac {d-1}{d}p)},
$$
$$
c_{m}=\frac {p}{d^{3}(1-\frac {d-1}{d}p)},\ 1\le m\le d-1.
$$
Applying Theorem 1 to each term $(\Phi _{k}\otimes Id)(y)$
included in the sum (\ref {Equat1}) we obtain the estimation
(\ref {XJ}) with $x_{j}^{s}=dTr((|f_{j}^{s}><f_{j}^{s}|\otimes
I_{K})x)$, where $|f_{j}^{s}>=W^{*}|e_{j}^{s}>,\ 0\le j,s\le d-1$.

$\Box $

Now consider the qubit case $d=2$. Let us involve the Pauli
operators such that
$$
U_{0,0}\equiv I_{H},\ U_{1,0}\equiv \sigma _{x},\ U_{0,1}\equiv
\sigma _{y},\ U_{1,1}\equiv i\sigma _{z}.
$$
Then, the maximum commutative groups ${\mathcal U}_{2}^{0}\equiv
{\mathcal U}_{y},\ {\mathcal U}_{2}^{1}\equiv {\mathcal U}_{z},\
{\mathcal U}_{2}^{2}\equiv {\mathcal U}_{x}$ are generated by the
spectral projections of the Pauli operators $\sigma _{y},\sigma
_{z}$ and $\sigma _{x}$, respectively. In the following we shall
also use the notation $e_{j}^{0}=e_{j}^{y},\ e_{j}^{1}=e_{j}^{z}$
and $e_{j}^{2}=e_{j}^{x},\ j=0,1,$ as well as $E_{0}=E_{y},\
E_{1}=E_{z},\ E_{2}=E_{x}$ for the corresponding conditional
expectations.

The "two-Pauli" channel
\begin {equation}\label {TP}
\Phi (\rho)=(1-2p)\rho+p\sigma _{y}\rho \sigma _{y}+p\sigma
_{z}\rho \sigma _{z},\ \rho \in \sigma (H),\ 0<p<\frac {1}{2}.
\end {equation}
is covariant with respect to ${\mathcal U}_{x}$ by means of
Proposition 6.

{\bf Proposition 8.}{\it The channel (\ref {TP}) can be
represented as a convex combination of the form
$$
\Phi (\rho)=\frac {1-3p}{1-p}\Phi _{1}(\rho)+\frac {2p}{1-p}\sigma
_{z}\Psi _{1}(\rho)\sigma _{z},
$$
where
$$
\Phi _{1}(\rho)=(1-p)\rho +p\sigma _{y}\rho \sigma _{y}
$$
is the phase damping and the channel
$$
\Psi _{1}(\rho)=p\rho +\frac {1-p}{2}\sigma _{x}\rho \sigma
_{x}+\frac {1-p}{2}\sigma _{z}\rho \sigma _{z}
$$
is covariant with respect to ${\mathcal U}_{y}$. }

{\bf Proof.}

It is straightforward to check the validity of the formula. To
show that the channel $\Psi _{1}$ is covariant with respect to
${\mathcal U}_{y}$ let us redefine the correspondence between the
Pauli operators and the Weyl operators such that
$$
\sigma _{y}\equiv U_{1,0},\ \sigma _{z}\equiv U_{0,1},\ \sigma
_{x}\equiv -iU_{1,1}.
$$
Then, the result follows from Proposition 6.

$\Box $

In the proof of Theorem 3 we shall need the "two Pauli" channel
of the form
\begin {equation}\label {TP2}
\Psi _{1}(\rho)=p\rho +\frac {1-p}{2}\sigma _{x}\rho \sigma
_{x}+\frac {1-p}{2}\sigma _{z}\rho \sigma _{z},\ \rho \in \sigma
(H).
\end {equation}

It is straightforward to check that the following estimation
holds.

{\bf Proposition 9.}{\it Suppose that $p\le \frac {1}{3}$. Then,
the channel (\ref {TP2}) is a convex combination of the form
$$
\Psi _{0}(\rho)=\frac {p}{1-p}\Phi _{0}(\rho)+\frac
{1-3p}{2(1-2p)}\sigma _{x}\Phi _{1}(\rho)\sigma _{x}+
$$
$$
(1-\frac {p}{1-p}-\frac {1-3p}{2(1-2p)})\sigma _{z}\Phi
_{1}(\rho)\sigma _{z},
$$
where
$$
\Phi _{0}(\rho)=(1-p)\rho +p\sigma _{x}\rho \sigma _{x},
$$
$$
\Phi _{1}(\rho)=(1-p)\rho +p\sigma _{y}\rho \sigma _{y},
$$
$\rho \in \sigma (H)$, are two phase damping channels. }

{\bf Proof of Theorem 3.}

Fix the state $\rho \in \sigma (H\otimes K)$. Due to Proposition
1 one can find the unitary operator $W\in {\mathcal U}_{x}$ such
that $Tr_{K}(\tilde \rho)=WTr_{K}(\rho)W^{*}$ satisfies the
identity
\begin {equation}\label {E1}
E_{y}(Tr_{K}(\tilde \rho))=\frac {1}{2}I_{H},
\end {equation}
where $\tilde \rho =(W\otimes I_{K})\rho (W^{*}\otimes I_{K})$.
Using the covariance of the channel $\Phi $ with respect to
${\mathcal U}_{x}$ we obtain
$$
S((\Phi \otimes Id)(\rho))=S((\Phi \otimes Id)(\tilde \rho)).
$$

It follows from Proposition 8 that
$$
(\Phi \otimes Id)(\tilde \rho)= \frac {1-3p}{1-p}(\Phi _{1}\otimes
Id) (\tilde \rho)+\frac {2p}{1-p}(\sigma _{z}\otimes I_{K})(\Psi
_{1}\otimes Id)(\tilde \rho )(\sigma _{z}\otimes I_{K}),
$$
where
$$
(\Phi _{1}\otimes Id)(\tilde \rho)=(1-p)\tilde \rho +p(\sigma
_{y}\otimes I_{K})\tilde \rho (\sigma _{y}\otimes I_{K}),
$$
and
$$
(\Psi _{1}\otimes Id)(\tilde \rho )=p\tilde \rho +\frac
{1-p}{2}(\sigma _{x}\otimes I_{K})\tilde \rho (\sigma _{x}\otimes
I_{K}) +\frac {1-p}{2}(\sigma _{z}\otimes I_{K})\tilde \rho
(\sigma _{z}\otimes I_{K}).
$$
Hence, we get
\begin {equation}\label {I1}
S((\Phi \otimes Id)(\tilde \rho))\ge \min \{S(\Phi _{1}\otimes
Id)(\tilde \rho),\ S((\Psi _{1} \otimes Id)(\tilde \rho ))\}.
\end {equation}

Taking into account (\ref {E1}) one can apply Theorem 1 to $(\Phi
_{1}\otimes Id)(\tilde \rho)$. Then,
$$
S((\Phi _{1}\otimes Id)(\tilde \rho))\ge -(1-p)\log (1-p)-p\log p+
\frac {1}{2}(S(2Tr_{H}((|e_{1}^{1}><e_{1}^{1}|\otimes I_{K})\rho
))+
$$
\begin {equation}\label {I2}
S(2Tr_{H}((|e_{2}^{1}><e_{2}^{1}|\otimes I_{K})\rho )),
\end {equation}
where $|e_{1}^{1}>=W^{*}|e_{0}^{y}>,\
|e_{2}^{1}>=W^{*}|e_{1}^{y}>$.

It follows from Proposition 1 that there exists the unitary
operator $\tilde W\in {\mathcal U}_{y}$ such that $Tr_{K}(\tilde
{\tilde \rho})=\tilde WTr_{K}(\tilde \rho )\tilde W^{*}$ satisfies
the identity
\begin {equation}\label {E2}
E_{x}(Tr_{K}(\tilde {\tilde \rho}))=\frac {1}{2}I_{H},
\end {equation}
where $\tilde {\tilde \rho} =(\tilde W\otimes I_{K})\tilde \rho
(\tilde W^{*}\otimes I_{K})$. Moreover,
\begin {equation}\label {E3}
E_{y}(Tr_{K}(\tilde {\tilde \rho}))=E_{y}(\tilde WTr_{K}(\tilde
\rho )\tilde W^{*})=E_{y}(Tr_{K}(\tilde \rho ))=\frac {1}{2}I_{H},
\end {equation}
because $\tilde W$ belongs to the algebra generated by the
projections $|e_{0}^{y}><e_{0}^{y}|,\ |e_{1}^{y}><e_{1}^{y}|$,
such that $E_{y}(\tilde W)=\tilde W$.

The covariance of $\Psi _{1}$ with respect to ${\mathcal U}_{y}$
gives us
$$
S((\Psi _{1}\otimes Id)(\tilde \rho))=S((\Psi _{1}\otimes
Id)(\tilde {\tilde \rho})).
$$
It follows from Proposition 9 that the channel $\Psi _{1}\otimes
Id$ can be represented in the form
\begin {equation}\label {I3}
(\Psi _{1}\otimes Id)(\tilde {\tilde \rho})= \frac {p}{1-p}(\Phi
_{0}\otimes Id)(\tilde {\tilde \rho})+\frac {1-3p}{2(1-2p)}(\sigma
_{x}\otimes I_{K})(\Phi _{1}\otimes Id)(\tilde {\tilde
\rho})(\sigma _{x}\otimes I_{K})+
\end {equation}
$$
(1-\frac {p}{1-p}-\frac {1-3p}{2(1-2p)})(\sigma _{z}\otimes
I_{K})(\Phi _{1}\otimes Id)(\tilde {\tilde \rho})(\sigma
_{z}\otimes I_{K}),
$$
where
$$
(\Phi _{0}\otimes Id)(\tilde {\tilde \rho})=(1-p)\tilde {\tilde
\rho} +p(\sigma _{x}\otimes I_{K})\tilde {\tilde\rho } (\sigma
_{x}\otimes I_{K}),
$$
$$
(\Phi _{1}\otimes Id)(\tilde {\tilde \rho})=(1-p)\tilde
{\tilde\rho } +p(\sigma _{y}\otimes I_{K})\tilde {\tilde \rho }
(\sigma _{y}\otimes I_{K}),
$$

Taking into account the identities (\ref {E2}) and (\ref {E3})
one can apply Theorem 1 to $(\Phi _{0}\otimes Id)(\tilde {\tilde
\rho})$ and $(\Phi _{1}\otimes Id)(\tilde {\tilde \rho})$. Thus,
we obtain
\begin {equation}\label {I4}
S((\Phi _{0}\otimes Id)(\tilde {\tilde \rho}))\ge -(1-p)\log
(1-p)-p\log p+\frac
{1}{2}(S(2Tr_{H}((|e_{1}^{2}><e_{1}^{2}|\otimes I_{K})\rho))+
\end {equation}
$$
S(2Tr_{H}((|e_{2}^{2}><e_{2}^{2}|\otimes I_{K})\rho ))),
$$
where $|e_{1}^{2}>=W^{*}\tilde W^{*}|e_{0}^{x}>,\
|e_{2}^{2}>=W^{*}\tilde W^{*}|e_{1}^{x}>$ and
\begin {equation}\label {I5}
S((\Phi _{1}\otimes Id)(\tilde {\tilde \rho}))\ge -(1-p)\log
(1-p)-p\log p+\frac
{1}{2}(S(2Tr_{H}((|e_{1}^{3}><e_{1}^{3}|\otimes I_{K})\rho))+
\end {equation}
$$
S(2Tr_{H}((|e_{2}^{3}><e_{2}^{3}|\otimes I_{K})\rho ))),
$$
where $|e_{1}^{3}>=W^{*}\tilde W^{*}|e_{0}^{y}>,\
|e_{2}^{3}>=W^{*}\tilde W^{*}|e_{1}^{y}>$. Combining (\ref {I1}),
(\ref {I2}), (\ref {I3}), (\ref {I4}) and (\ref {I5}) we obtain
the result.

$\Box $

\section*{Acknowledgments} The author is grateful to Mary Beth Ruskai for the
remark about the "two-Pauli" channel and sending a draft of her
unpublished manuscript which partially inspired this work. This
paper would not be possible without a detailed report on its
first version by the anonymous referee I like to thank especially.

\end {document}